\documentclass[12pt,a4paper,final]{iopart}

\usepackage{iopams}
\expandafter\let\csname equation*\endcsname\relax
\expandafter\let\csname endequation*\endcsname\relax
\usepackage{amsmath}

\usepackage[T1]{fontenc}
\usepackage[colorlinks=true]{hyperref}
\usepackage{amssymb}
\usepackage{graphicx}
\usepackage{color}
\usepackage{subfig}
\usepackage{caption}
\usepackage{placeins}
\usepackage{cleveref}

\usepackage{ulem} 

\begin{document}

\title[Low-latency analysis pipeline for CBCs in the advanced GW detector era]{Low-latency analysis pipeline for compact binary coalescences in the advanced gravitational wave detector era}

\author{T.~Adams$^{1}$, D.~Buskulic$^{1}$, V.~Germain$^{1}$, G.~M.~Guidi$^{2,3}$, F.~Marion$^{1}$, M.~Montani$^{2,3}$, B.~Mours$^{1}$, F.~Piergiovanni$^{2,3}$, G.~Wang$^{2,4}$}

\address{$^{1}$ Laboratoire d'Annecy-le-Vieux de Physique des Particules (LAPP), Universit\'e de Savoie, CNRS/IN2P3, F-74941 Annecy-le-Vieux, France }
\address{$^{2}$ Universit\`a degli Studi di Urbino `Carlo Bo', I-61029 Urbino, Italy }
\address{$^{3}$ INFN, Sezione di Firenze, I-50019 Sesto Fiorentino, Firenze, Italy }
\address{$^{4}$ INFN, Gran Sasso Science Institute, I-67100 L'Aquila, Italy }

\ead{Thomas.Adams@lapp.in2p3.fr}

\begin{abstract}
The Multi-Band Template Analysis (MBTA) pipeline is a low-latency coincident analysis pipeline for the detection of gravitational waves (GWs) from compact binary coalescences.
MBTA runs with a low computational cost, and can identify candidate GW events online with a sub-minute latency.
The low computational running cost of MBTA also makes it useful for data quality studies.
Events detected by MBTA online can be used to alert astronomical partners for electromagnetic follow-up.
We outline the current status of MBTA and give details of recent pipeline upgrades and validation tests that were performed in preparation for the first advanced detector observing period.
The MBTA pipeline is ready for the outset of the advanced detector era and the exciting prospects it will bring.
\end{abstract}

\section{Introduction}

At the start of the advanced detector era, the LIGO \cite{Abbott:2009li} and Virgo \cite{Accadia:2012vi} detectors are restarting observation after being out of operation for several years while they underwent a period of upgrades and commissioning.
In 2015, the two Advanced LIGO \cite{Aasi:2015al} detectors begin the first coincident observing period of the advanced detector era, O1.
The Advanced Virgo detector \cite{Acernese:2015va} is expected to come online in 2016 to take part in the second observing period, O2 \cite{Aasi:2013lo}.
In the lead up to each observing period are a number of engineering runs which are used to test the detector and analysis pipeline infrastructure as the detectors are recommissioned and brought into an observational state.

Some of the most well understood and strongest potential sources of GWs for ground based GW detectors are compact binary coalescences (CBCs).
These consist of binary neutron stars (BNS), a neutron star and a black hole (NSBH), or binary black holes (BBH).
It is possible to model templates of the GW signals that these sources will produce and perform a matched filtered analysis.
BNS and NSBH mergers both have many possible mechanisms for producing EM counter parts \cite{Metzger:2011wa}, which make them interesting sources for GW-EM multi-messenger astronomy \cite{Aasi:2013lo}.
The BNS detection rate of $2 \times 10^{-4} - 0.6\,\textrm{yr}^{-1}$ in the initial detector era will improve to $0.4 - 1000\,\textrm{yr}^{-1}$ in the advanced detector era at the design sensitivity \cite{Abadie:2010fn}.

Running an online analysis presents a number of challenges such as the availability of minimal data quality information to veto instrumental and environmental noise, and the requirement to analyse data faster than real time to maintain an  online analysis.
The Multi-Band Template Analysis (MBTA) \cite{Abadie:2012wx,Adams:2015mb} is a low-latency, computationally cost effective, coincidence analysis pipeline used to detect GWs from CBCs.
MBTA uses the standard matched filter \cite{Waistein:1962vz} to extract CBC signals from the GW channel data of each detector in the network independently, before results are combined to find GW candidate events.
The focus of MBTA is the online detection of GW candidate events with sub-minute latency, but it can also be used for data quality studies due to its low computational cost.
MBTA determines GW candidate event significance by calculating the false alarm rate (FAR) using data immediately before the event to evaluate the detector background noise at the time of the event.

\begin{figure}
  \centering
  \includegraphics[width=0.7\textwidth]{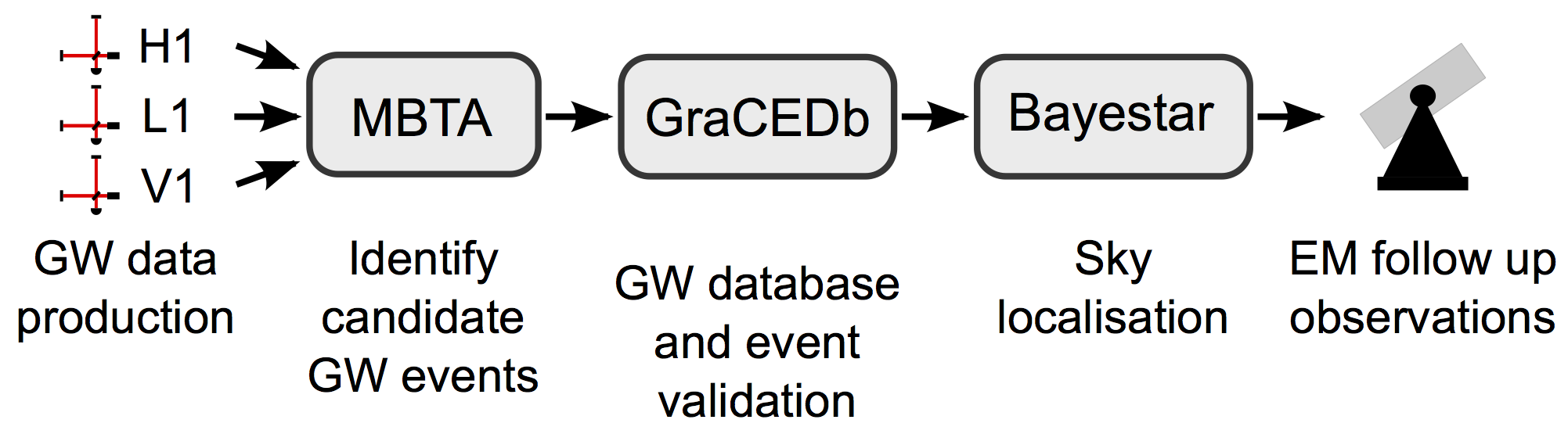}
  \caption{\label{fig:analysistimeline}Overview of the GW-EM follow-up pipeline.
           Data from the GW detectors is processed by MBTA, which identifies GW candidate events.
           These events are uploaded to GraCEDb which triggers further event validation and sky localisation with Bayestar.
           Events which pass all validation tests are released to astronomical partners for EM follow-up.}
\end{figure}

GW candidate events detected by MBTA are uploaded to the Gravitational Wave Candidate Event Database (GraCEDb) \cite{gracedb}, an automated archive where details about the GW candidate event and follow-up studies are recorded.
There are a number of other modelled GW search pipelines which also upload events to GraCEDb, namely GSTLAL \cite{Privitera:2014gs} and pyCBC \cite{Usman:2015tc}, as well as unmodelled search pipelines such as CWB \cite{Klimenko:2008ts}.
MBTA events uploaded to GraCEDb are analysed with Bayestar \cite{Singer:2014iq}, a rapid Bayesian position reconstruction code that produces probability sky maps for the sky localisation of GW candidate events.
GW candidate events with high significance are validated by a number of human monitors.
Events which pass the validation process are distributed as a LIGO/Virgo GCN CIRCULAR \cite{lvcgcncircular} to astronomical partners for EM follow-up.
An overview of the GW-EM follow-up pipeline including MBTA is given in figure\,\ref{fig:analysistimeline}.

In this paper the key elements of the MBTA pipeline are outlined in section\,\ref{sec:mbta}.
In section\,\ref{sec:improvements} details are given for recent pipeline upgrades and pipeline validation tests that were performed in preparation for the advanced detector era.
Finally, in section\,\ref{sec:adv_era} the status of MBTA is summarised for operations in the first observing period of the advanced detector era and future developments are outlined.

\section{The Multi-Band Template Analysis}
\label{sec:mbta}

\begin{figure}
  \centering
  \includegraphics[width=0.95\textwidth]{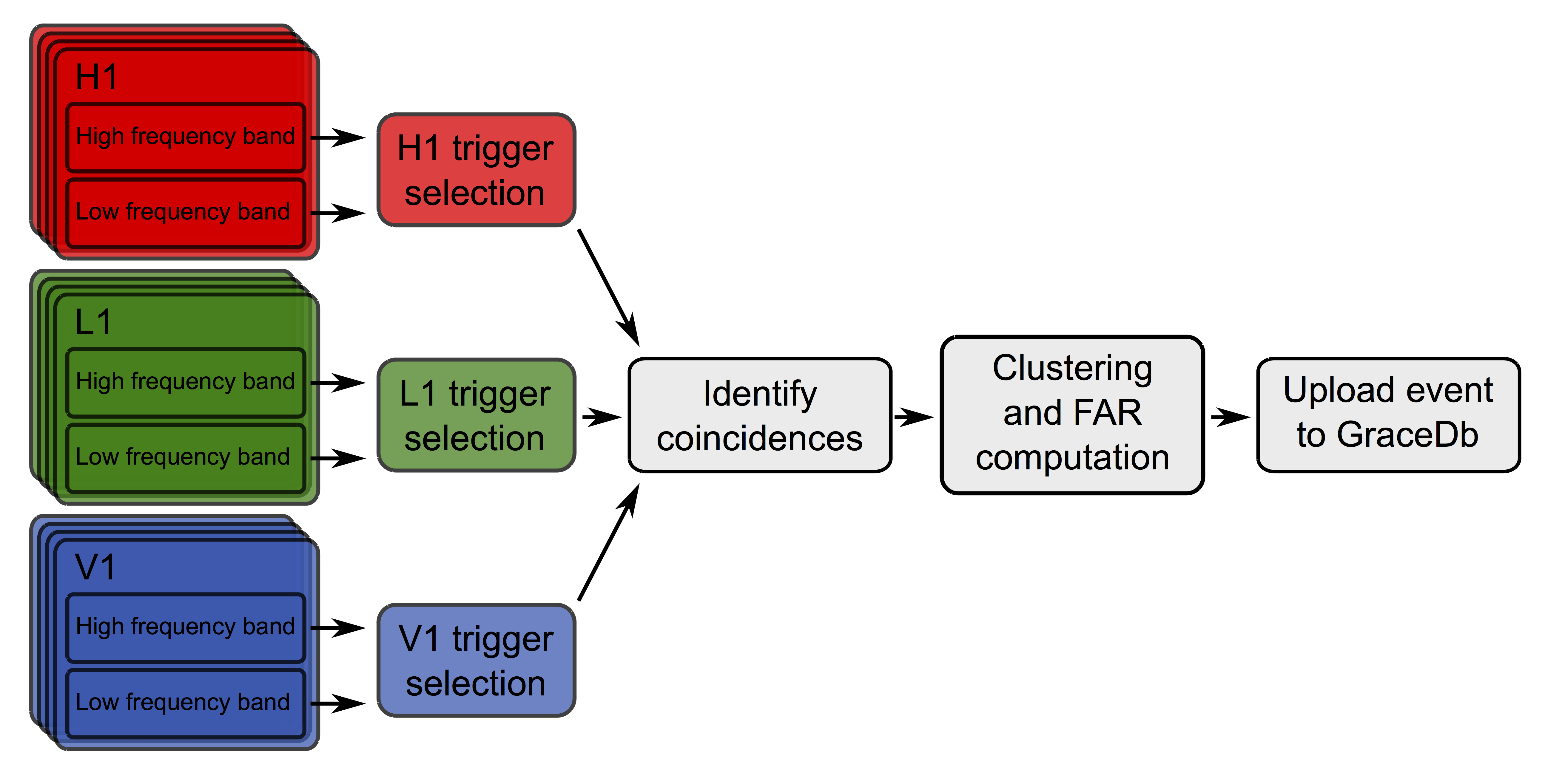}
  \caption{\label{fig:mbtatimeline}Overview of the MBTA pipeline.
           The multi-band matched filter is performed, split across multiple jobs for each detector.
           These results are then combined to produce the individual detector trigger lists which are subjected to trigger selection by the $\chi^{2}$ cut, matched filter time-series signal consistency test, and data quality checks.
           Coincident triggers are identified between detectors, using timing and the exact match coincidence test, the FARs are computed and triggers are clustered to produce coincident events.
           Coincident events with low FAR below some configurable threshold are uploaded to GraCEDb to be validated as GW candidate events.}
\end{figure}

In this section the main elements of the MBTA pipeline, GW candidate event identification, and GW alert generation for EM follow-up will be outlined.
An overview of the MBTA pipeline is given in figure\,\ref{fig:mbtatimeline}.

\subsection{Single detector analysis}
\label{sec:single_det}

The MBTA pipeline performs a coincident analysis, analysing each detector in the network separately before the results are combined to identify coincident events.
To reduce the computational cost of the matched filtering, which is the most expensive element of the analysis, MBTA uses the novel approach of splitting the matched filter across two (or more) frequency bands.
The boundary frequency between the low frequency (LF) and high frequency (HF) bands, $f_{\mathrm{c}}$, is selected so that the signal-to-noise ratio (SNR) is shared roughly equally between the low and high frequency bands, typically $f_{\mathrm{c}} \approx 100$\,Hz, for the expected sensitivity curve of the advanced detectors.
The analysis bandwidth for the advanced detector era will typically be $30$\,Hz to $2048$\,Hz.
This multi-band analysis procedure gives a reduction in the computational cost, while losing no SNR on average compared to a matched filter performed with a single band analysis \cite{Marion:2003vp}.
Recent studies have found a reduction in the computational cost of the online pipeline configuration relative to an equivalent single band analysis of a factor of $\sim7$, while the offline pipeline configuration gives a larger reduction of a factor of $\sim11$.

The reduction in the computational cost is achieved by using shorter templates in each frequency band, and so the phase of the signal is tracked over fewer cycles.
This reduces the number of templates that are required to cover the equivalent parameter space of a single band analysis.
Another benefit of using a multi-band analysis is that a reduced sampling rate in time for the low frequency band can be used, by down-sampling in the frequency domain, which reduces the cost of the Fast Fourier Transforms (FFTs) involved in the filtering.

Before running MBTA online, the pipeline must be initialised and a fixed bank of inspiral templates is constructed with a typical minimal match of $97\%$ \cite{Brown:2012gz}.
In preparation for the advanced detector era, MBTA is now able to run using aligned spin template banks which model the effect of the binary component spins on the waveform.
Currently MBTA uses SpinTaylorT4 waveforms, which are post-Newtonian inspiral time-domain waveforms that allow for misaligned precessing component spins \cite{Buonanno:2003wf}.
Different choices of the waveform could affect the search efficiency \cite{Canton:2014kg}; in fact there are systematic differences between waveform approximants which could impact the matched filter SNR computation \cite{Buonanno:2009gk} \cite{Nitz:2013:mo}.
However, this will not impact seriously the results presented here and it will be studied in detail elsewhere.
An overview of the performance of the analysis using an aligned spin template bank, and validation of the pipeline is given in section\,\ref{sec:spintemplates}.
The waveform generation and geometric-based template placement is performed using the LALsuite \cite{lalsuite}.
This template bank, which covers the parameter space of interest for the specific analysis, is generated at initialisation using a reference noise power spectrum taken at a time when the detectors are performing well.
In O1, BNS and NSBH template bank will be used as described in section\,\ref{sec:signalrecovery}.

This template bank is referred to as the ``virtual'' template bank, as it is not actually used to perform the matched filtering, as explained below.

Running the multi-band matched filtering on each frequency band requires a separate ``real'' template bank for each frequency band, which is actually used in the matched filtering of the data.
The waveforms and the template placement method for the real template banks are the same used for the virtual template bank, but with the $97\%$ minimal match computed over the relevant frequency range.
During initialisation each virtual template is associated with a real template in each frequency band.
To perform this association, real templates in each band are match filtered with the virtual template to find the real templates in each band which have the maximal match.
The combination parameters, $\Delta t$ and $\Delta \phi$ which are used when performing the coherent sum of the multi-band results, are also determined from the difference in time and phase between the real templates in the different frequency bands.
The filtering produces the matched filter time-series for each frequency band, both in-phase and in-quadrature.

MBTA requires the virtual template bank, real template banks for each frequency band, calibrated GW channel data from each detector in the network, as well as any available data quality information to perform an analysis online.
The power spectral density (PSD) used for the matched filtering is updated using a running average of data that is deemed of observational quality, passing data quality tests.
An overview of the online configuration tests is given in section\,\ref{sec:online}.

To combine the multiple frequency bands, the matched filter output is examined in an iterative way.
The maximum of the SNR in the match filter output of each frequency band for a real template is compared with a threshold, typically $\textrm{SNR} > 5/\sqrt{2}$.
If the threshold is exceeded, combinations are made for all virtual templates associated to this real template.
The low frequency band SNR time series is up-sampled with a quadratic interpolation to the sampling frequency of the high frequency band.

The complex output of the match filter for the virtual template (VT) is constructed from the output of the match filter of the low frequency (LF) and high frequency (HF) real templates (RT), using the combination parameters $\Delta t$ and $\Delta \phi$ (time and phase offset between the LF and HF bands):
\begin{equation}
    \left\langle h, \mathrm{VT} \right\rangle (t) = \left\langle h, \mathrm{RT_{LF}} \right\rangle(t) + e^{i \Delta \phi} \left\langle h, \mathrm{RT_{HF}} \right\rangle(t + \Delta t) \, ,
\end{equation}
where
\begin{equation}
    \left\langle h, \mathrm{T} \right\rangle (t) = 4 \mathcal{R} \int_{\Delta f} \frac{\tilde{h}(f)\tilde{T}^{*}(f)}{S_{h}(f)} e^{-2 i \pi f t} \delta f \, ,
\end{equation}
using a sampling frequency for the match filter in each frequency band of twice the upper frequency cut-off for that band.

The modulus of the match filter output is then examined at its maximum value to extract the signal parameters (time of arrival, coalescence phase, full band SNR).
The full band SNR is compared to a global threshold, typically SNR$ > 5$, to produce the single detector trigger list.

A simple benefit of the multi-band analysis procedure is the availability of a computationally inexpensive $\chi^{2}$ calculation \cite{allen:2005ch}.
This can be used to perform a signal consistency test, the $\chi^{2}$ cut, which checks that the partition of the SNR of a trigger between the multiple frequency bands is consistent with what is expected for a true signal.
For the two band analysis the $\chi^{2}$ cut is defined as
\begin{equation}
    \chi^{2} < \alpha (2 + \beta \times \textrm{SNR}^{2}) \, ,
\end{equation}
with the empirical values $\alpha = 3$ and $\beta = 0.025$.

\begin{figure}
  \centering
  \includegraphics[width=0.7\textwidth]{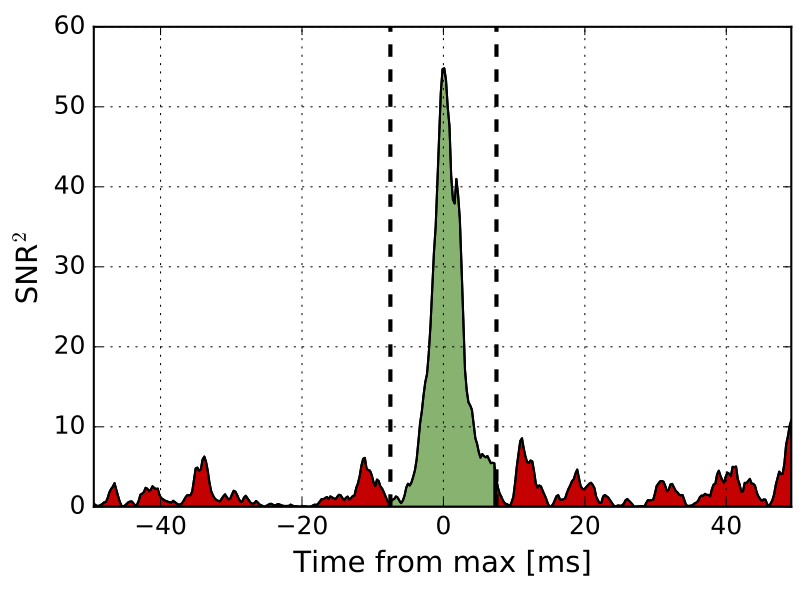}
  \caption{\label{fig:mfo} $\mathrm{SNR}^{2}$ time series from the matched filter output for a simulated GW in the presence of noise.
           The central window around the maximum (dashed lines) is $\pm 7.5$\,ms while the total time is $0.1$\,s.
           The ratio of the integrated $\mathrm{SNR}^{2}$ inside the central window (light green) to the integrated $\mathrm{SNR}^{2}$ in the surrounding data (dark red) can be used to veto events that do not behave like GW signals.}
\end{figure}

During pipeline testing in preparation for the advanced detector era, it was found that loud noise events were contaminating the background used for the FAR estimation of simulated signals added to the data for testing the pipeline, injections.
The $\chi^{2}$ cut alone was not sufficient to reject all of these noise events, and so a new signal based consistency test was added to MBTA, the matched filter time-series signal consistency test.
The matched filter time-series for a GW signal will have a single narrow peak, whereas for noise the matched filter time-series will have a much broader peak with multiple maximums around the central feature \cite{Guidi:2004ve,Shawhan:2004ve, Hanna:2008vj}.
Finding the ratio $R$ of the integrated $\mathrm{SNR}^{2}$ in the $0.1$\,s surrounding a trigger excluding a small central window of $\pm 7.5$\,ms, to the integrated $\mathrm{SNR}^{2}$ inside this central window, triggers that do not behave like signals can be rejected.
The windowing use in the matched filter time-series signal consistency test for a simulated GW can be seen in figure \,\ref{fig:mfo}.
Taking the integrals on the $\mathrm{SNR}^{2}$, triggers with
\begin{equation} \label{eq:mfo}
    R > A/\mathrm{SNR}^{2}_{peak} + B
\end{equation}
are rejected.
$\mathrm{SNR}^{2}_{peak}$ is the squared peak SNR of the trigger and the typical empirical values $A = 65$ and $B = 0.4$ are used.
More details and validation of the signal based consistency test can be found in section\,\ref{sec:mfo}.

\subsection{Identify coincidences events}

The single detector trigger lists are combined to find coincidence events between two (or more) GW detectors and the combined SNR is computed.
Coincidence events are identified by MBTA using time coincidence and the exact match coincidence test.
The time coincidence of events is checked between detectors, taking into account both the time of flight between the detectors and the experimental uncertainty in the arrival time measurement.
The exact match coincidence test requires that events are found in all detectors with the same template parameters; component masses and spins \cite{Usman:2015tc}.
To use exact match coincidence test the same virtual template banks must be used for each detector.
This procedure gives improved background rejection, compared to the chirp mass consistency test which was previously used in MBTA, as well as better timing accuracy between detectors at the price of higher computational cost.
The exact match coincidence test is expected to remain the default for the advanced detector era.

The significance of each coincident trigger is estimated by calculating the FAR, the expected rate of coincidence triggers from noise only (glitches) that have an equal or larger combined SNR than the coincident trigger.
The FAR computation is based on the assumption that the detectors are independent, producing uncorrelated triggers.
The average single detector background trigger rate $R_{i}$ is simply $N_{i}/T_{i}$, where $N_{i}$ is the number of background triggers accumulated in a given time window $T_{i}$ immediately before the coincident trigger for each detector in the network ``i, j, k''.

The FAR estimation for coincidence between two (or more) detectors is split into two parts:
\begin{itemize}
    \item \textbf{The effect of the time coincidence:}
    Assuming stationary single detector background trigger rates $R_{i}$ and $R_{j}$ (and $R_{k}$ for a triple coincidence), and a coincidence time window $\delta t_{ij}$ (and $\delta t_{ik}$ for a triple coincidence), the rate of random coincidence is given by the single detector background trigger rates multiplied by the coincidence time window $R_{i} R_{j} \delta t_{ij}$ (or $R_{i} R_{j} \delta t_{ij} R_{k} \delta t_{ik}$ for a triple coincidence).
    \item \textbf{The effect of the parameter match due to the exact match coincidence test:}
    This is measured by using the most recent $N_{i}$ triggers in the background time window $T_{i}$ for each detector, making the $N_{i} N_{j}$ combinations independent of their arrival time (or $N_{i} N_{j} N_{k}$ for a triple coincidence).
    Finding the number $M$ of these background coincidences that pass the exact match coincidence test and have a combined SNR larger the coincident trigger.
    The probability to pass the test is simply $M/(N_{i} N_{j})$ (or $M/(N_{i} N_{j} N_{k})$ for a triple coincidence).
\end{itemize}
Therefore for a double coincidence trigger,
\begin{align}
\begin{split}
    \mathrm{FAR_{double}} &= R_{i} R_{j} \delta t_{ij} \frac{M}{(N_{i} N_{j})} \, ,\\
                 &= \frac{M \delta t_{ij}}{T_{i} T_{j}} \, ,
\end{split}
\end{align}
and for a triple coincidence trigger,
\begin{align}
\begin{split}
    \mathrm{FAR_{triple}} &= R_{i} R_{j} \delta t_{ij} R_{k} \delta t_{ik} \frac{M}{(N_{i} N_{j} N_{k})} \, ,\\
                 &= \frac{M \delta t_{ij} \delta t_{ik}}{T_{i} T_{j} T_{k}} \, .
\end{split}
\end{align}

The coincidence time window, $\delta t_{ij}$, is set to $15$\,ms, accounting for the time of flight between the detectors and the experimental uncertainty in the arrival time measurement for the HL network.
To reach small FARs the background time window, $T_{i}$ can be increased.
However, the purpose of this window is to sample the detector background in the time close to the coincidence trigger, so it is not practical to increase this number to produce arbitrarily small FARs.
A value of $T_{i} = 3$\,hours was selected as a period over which the detectors can be considered reasonably stable.
This allows a minimum achievable FAR of $\mathtt{\sim}1\times10^{-10}$\,Hz (once per $300$\,years).
Much smaller FARs can be obtained with ``deeper'' offline searches but the goal for MBTA is the online search for GW.

When calculating the FAR for loud triggers, there can be ``satellite'' triggers around the central trigger which must be removed from the FAR computation to avoid biasing it.
This is done by excluding triggers in a veto window around a loud trigger, including the trigger itself.
To study the possibility of a bias in the FAR calculation due to this exclusion, the FAR without applying the veto window is also calculated.

When analysing data from three detectors, there are effectively four searches being performed.
There are the three double coincidence searches (HL, HV, LV) and a triple coincidence search (HLV).
To take into account the trials factor, the FAR for coincident triggers detected during triple coincident time must be multiplied by 4.

After coincidence triggers are identified clustering is performed to produce events.
When finding the FAR for an event, the number of coincidence triggers that were clustered into the event must taken into account and applied as a scaling factor for the FAR.
The average number of coincidence triggers clustered is computed for the most recent coincident events, typically the last $200$, excluding events with a large number of triggers in order to avoid bias due to loud instrumental glitches.
This is then applied as a scaling factor, typically having a value of $\mathtt{\sim}3$, when calculating the FAR for the event.

\subsection{Alert and EM follow-up}

\begin{figure}
  \centering
  \includegraphics[width=0.7\textwidth]{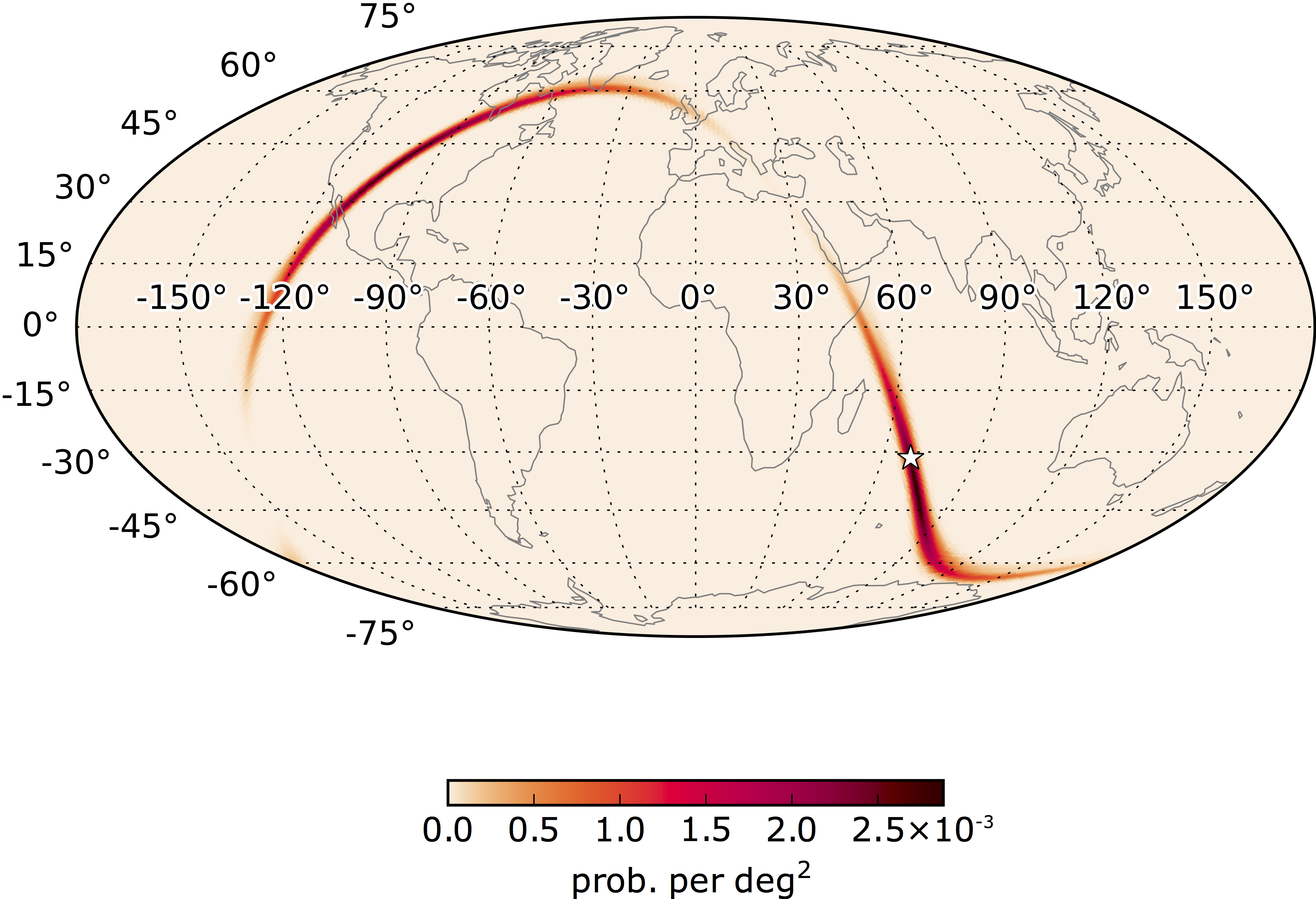}
  \caption{\label{fig:skymap}Localisation skymap produced with Bayestar for a simulated GW event detected with the HL network by MBTA in the sky localisation MDC.
           A Mollweide projection in geographic coordinates is used.
           The star shows the sky location of the injected GW signal and the shading is proportional to posterior probability per deg$^2$.}
\end{figure}

Low FAR events produced by MBTA which pass a configurable FAR threshold are submitted to GraCEDb.
This threshold is selected to allow study of the FAR distribution of events produced by MBTA, whilst not over burdening GraCEDb.
A typical threshold is $1\times10^{-4}$\,Hz (once per $\mathtt{\sim}3$\,hours).
MBTA uploads information about the event, as well as a PSD for each detector at the time of the event.
The time, amplitude, and phase information reported by MBTA are used by Bayestar to produce a probability sky map that is appended to the GraCEDb event which can be used to plan EM follow-up observations.
An example sky map is shown in figure\,\ref{fig:skymap}.
For validation of the sky localisation performance of MBTA see section\,\ref{sec:exactmatch}.
MBTA also uploads a number of plots which demonstrate the pipeline performance around the time of the event.
Currently these are matched filter time-series for each of the detectors, SNR time-series, trigger rate time-series, and chirp mass against time plots.
The behaviour of these plots is understood in the presence of both signal and noise events, which makes them useful during GW candidate event follow-up and validation.

\section{Pipeline validation}
\label{sec:improvements}

In this section details of the pipeline validation tests that were performed for a number of recent upgrades to the pipeline will be given.
These upgrades were performed in the build up to the advanced detector era, and were tested with a number of mock data challenges (MDCs) and engineering runs.

One of the MDCs that was used for many of the following investigations is the ``BNS MDC''.
The data used in this MDC is initial LIGO data which has been recoloured to match the expected PSD of the early Advanced LIGO detectors during O1.
This MDC contains 3.5PN SpinTaylorT4 BNS injections with component masses of $1\,\mathrm{M}_{\odot}$ to $3\,\mathrm{M}_{\odot}$ and dimensionless component spins of up to $0.4$ aligned with the orbital angular momentum.
The analysis duration was one million seconds, with an injection every $140$\,s randomly distributed over all sky positions and orientations, and uniformly distributed in chirp distance, defined for some fiducial chirp mass ($\mathcal{M}_{o}$) as,
\begin{equation}
  D_{chirp} = D_{\mathrm{eff}}(\mathcal{M}_{o}/\mathcal{M})^{5/6} \, .
\end{equation}
where $D_{\mathrm{eff}}$ is the effective distance which takes into account both the physical distance and orientation of the source \cite{Thorne:1987gr}.

\subsection{Online testing and MDCs}
\label{sec:online}

In preparation for running MBTA online during the advanced detector era, data which was streamed during the engineering runs to simulate the situation during O1 was analysed.
MBTA was run on this data to test the computing infrastructure used by the pipeline and to tune the online configurations.
There are a number of configuration options available in the pipeline, particularly for the matched filtering which is the most computationally expensive element of the pipeline, which allows tuning of the latency of the MBTA pipeline to some extent with the computing resources available.
For the online runs the MBTA pipeline is tuned to achieve sub-minute latency, which gives a good compromise between cost and latency.

One way to reduce the latency of the matched filtering in the analysis, at the expense of increased computational cost, is the optimise latency option.
Typically, FFTs in the low and high frequency band are performed with some configurable overlap between subsequent FFTs.
This is usually determined by the length of the templates in each frequency band.
However, because the low frequency band requires longer templates there is a disparity in the latency of the filtering for the multiple frequency bands.
Typically, the analysis latency is dominated by the low frequency band latency.
The optimise latency option helps to address this by increasing the overlap of the low frequency band FFTs to match the rate of the high frequency band FFTs, which are shorter and therefore performed more frequently.
With this option, the results for both frequency bands are produced simultaneously and the analysis is then dominated by the high frequency band latency, which is less due to the shorter templates.

Due to the improved low frequency sensitivity of the Advanced LIGO detectors, compared to the initial LIGO detectors, it is important to utilise this low frequency band to improve the SNR of signals that MBTA recovers.
For this the low frequency cut-off of the low frequency band was reduced from $50$\,Hz to $30$\,Hz.
Increasing the analysis bandwidth also increases the computational cost of the analysis, as searching at lower frequencies required the use of longer and more numerous templates.
However, due to the optimize latency option the low latency of the analysis can be maintained and the option to further reduce the low frequency cut-off remains for future development.

\subsection{Spinning templates}
\label{sec:spintemplates}

In preparation for searches in the advanced detector era, MBTA is now able to run with aligned spin template banks.
The generation of this kind of template bank has not been implemented as a MBTA functionality, but the aligned spin template banks are generated by the LALsuite \cite{Buonanno:2003wf, lalsuite} and then loaded into MBTA.
To test the performance of MBTA with aligned spin templates an investigation using the BNS MDC was performed.
Two parallel analyses were run on this MDC, one with non-spinning template banks and the other with aligned spin template banks \cite{Brown:2012gz,Canton:2014kg}, covering the same parameter space as the injections.

\begin{figure}
  \centering
  \includegraphics[width=0.7\textwidth]{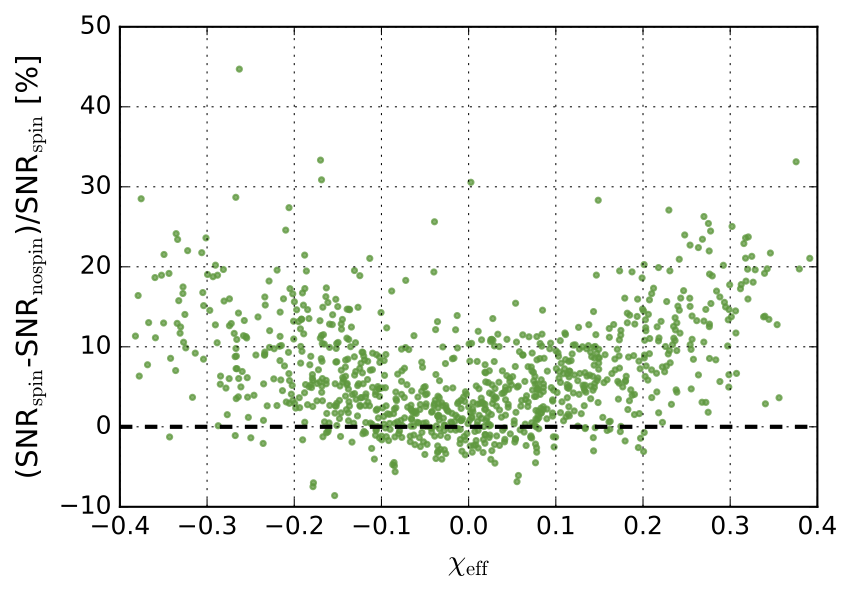}
  \caption{\label{fig:snr_scatter}The percentage improvement from using aligned spin templates compared to non-spinning templates against the effective dimensionless spin of injections in the BNS MDC.}
\end{figure}

\begin{figure}
  \centering
  \includegraphics[width=0.7\textwidth]{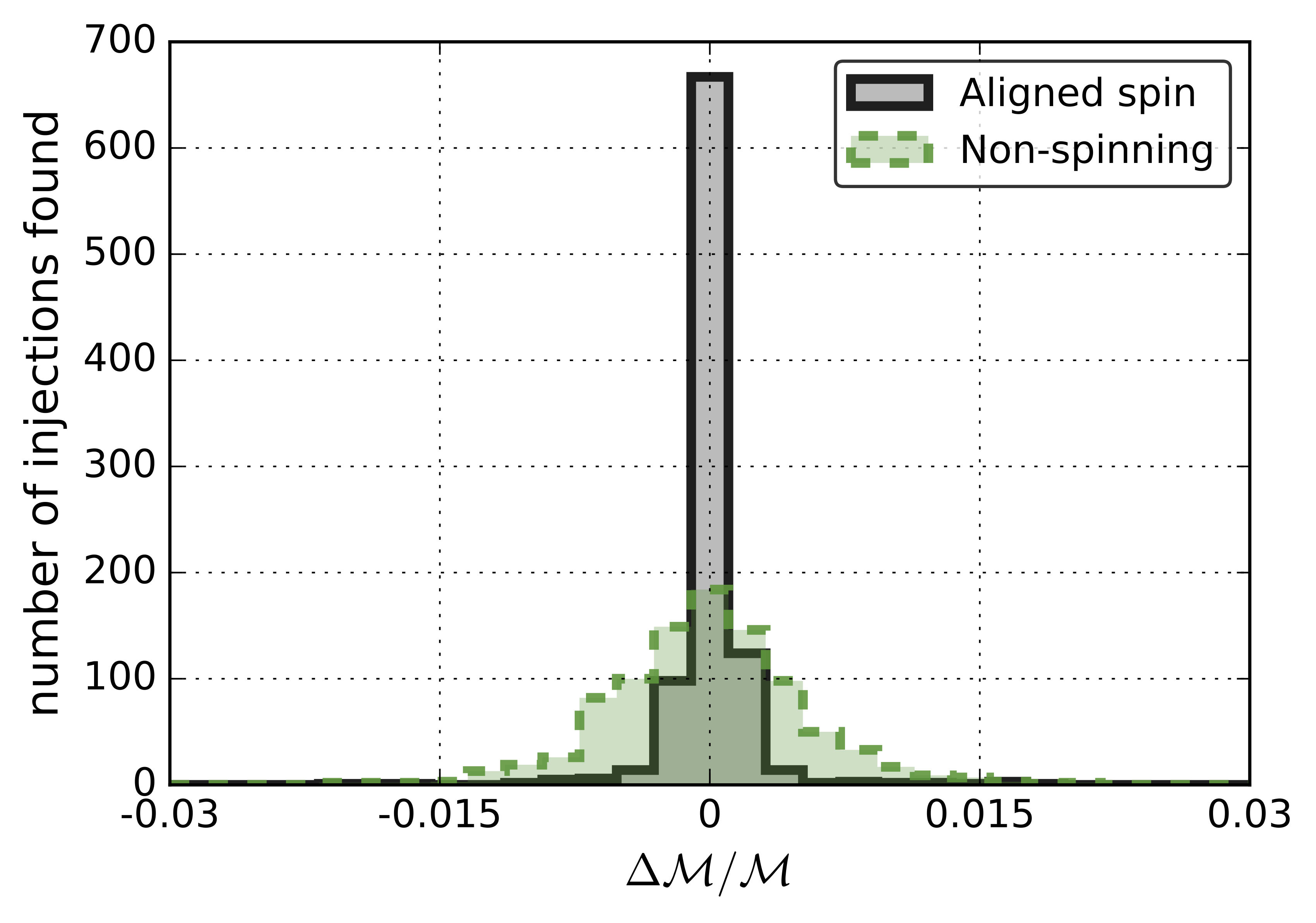}
  \caption{\label{fig:spin}Chirp mass ($\mathcal{M}$) reconstruction for the aligned spin and non-spinning template banks analyses of the BNS MDC.}
\end{figure}

\begin{figure}
  \centering
  \includegraphics[width=0.7\textwidth]{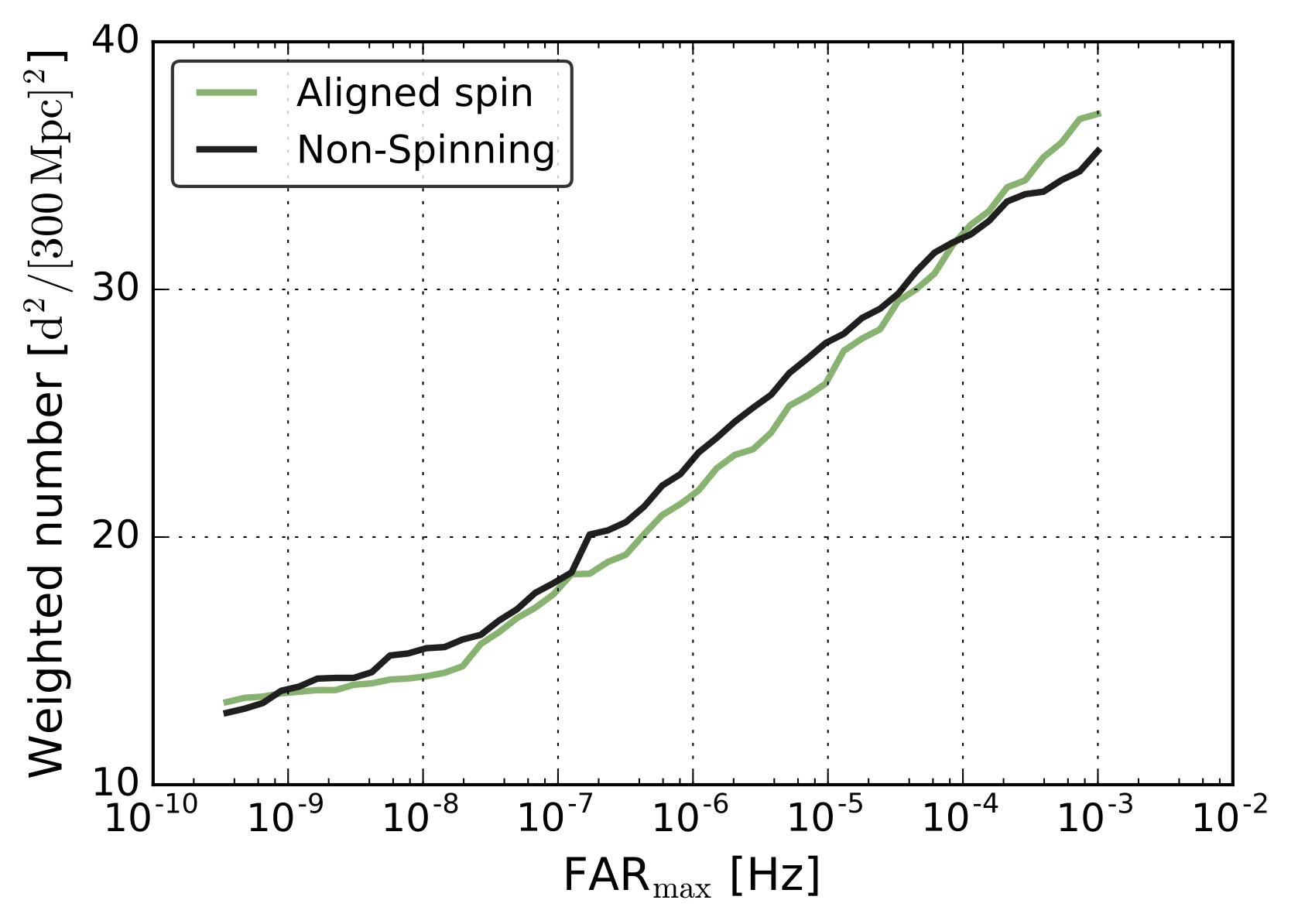}
  \caption{\label{fig:far_spin}Number of injections weighted by $d^{2}$ detected with FAR below FAR$_{\mathrm{max}}$ against FAR$_{\mathrm{max}}$ for the aligned spin and non-spinning template bank analyses of the BNS MDC.}
\end{figure}

As expected, using the aligned spin template bank gave an improvement of the reconstructed SNR for the commonly recovered injections.
This was particularly true for injections with large absolute values of the effective dimensionless spin $\chi_{\mathrm{eff}}$, defined as the weighted average of the two dimensionless spins,
\begin{equation}
    \chi_{\mathrm{eff}} = \frac{(\chi_{1} m_{1} + \chi_{2} m_{2})}{(m_{1} + m_{2})} \, .
\end{equation}
This can be seen in figure\,\ref{fig:snr_scatter}.
An improvement in the reconstructed chirp mass ($\mathcal{M}$) is obtained, shown by the narrower distribution compared to the non-spinning template analysis as seen in figure\,\ref{fig:spin}.
A modest improvement in the timing reconstruction, which is important for accurate sky localisation was also observed.
At a given SNR, the FAR for the aligned spin template bank analysis is larger than for the non-spinning template bank due to the aligned spin template bank containing around ten times more templates.
This results in slight loss of efficiency for the aligned spin compared to the non-spinning template bank analyses at low FAR values, as seen in figure\,\ref{fig:far_spin}.
However, as the actual spin distribution of sources in nature is not known and an improvement in the recovered SNR, chirp mass and timing reconstruction is observed, the aligned spin template bank will be used in future analyses.

\subsection{Signal recovery}
\label{sec:signalrecovery}

After the promising results of the BNS MDC investigations, a template bank which covers the parameter space of interest for EM follow-up was generated.
This template bank includes both BNS and NSBH templates with component masses ($m1, m2$) of $1\,\mathrm{M}_{\odot}$ to $12\,\mathrm{M}_{\odot}$ with $\mathcal{M} < 5\,\mathrm{M}_{\odot}$.
Components with a mass $<2\,\mathrm{M}_{\odot}$ have a dimensionless spin of up to $0.05$ and components with a mass $\geq 2\,\mathrm{M}_{\odot}$ have a dimensionless spin of up to $1$.

\begin{figure}
  \centering
  \includegraphics[width=0.7\textwidth]{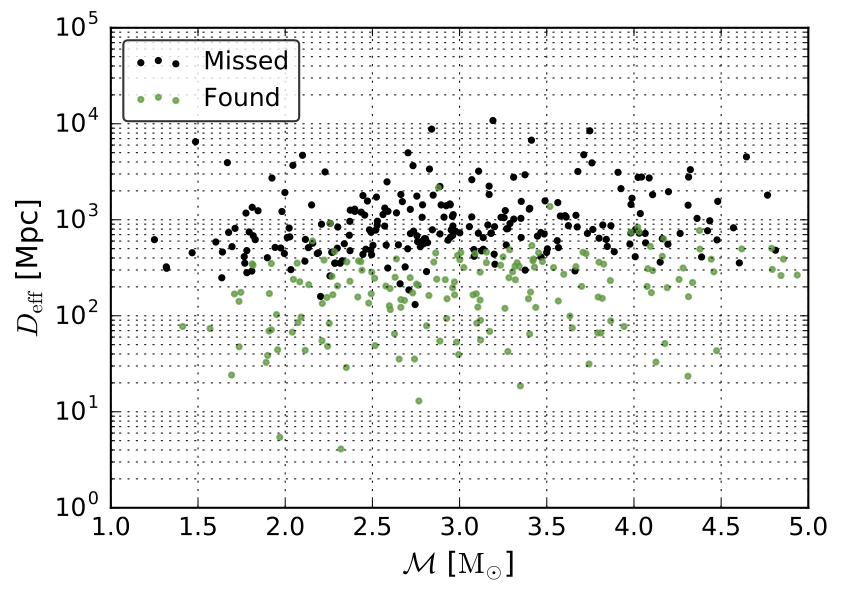}
  \caption{\label{fig:missedfound}Chirp mass ($\mathcal{M}$) against effective distance to demonstrate the signal recovery of injections by MBTA.
           Found injections are shown in light green while missed injections are shown in black.
           Here MBTA uses a BNS and NSBH template bank covering the parameters space of interest for EM follow-up to recover NSBH injections.}
\end{figure}

Using this BNS and NSBH template bank, the performance of MBTA to recover NSBH injections was studied with the ``NSBH MDC''.
This MDC uses initial LIGO data which has been recoloured to match the expected PSD of the early Advanced LIGO detectors during O1.
This MDC contains SpinTaylorT2 NSBH injections with NS of mass $1\,\mathrm{M}_{\odot}$ to $3\,\mathrm{M}_{\odot}$ with dimensionless component spins of up to $0.05$ and BH of mass $2\,\mathrm{M}_{\odot}$ to $12\,\mathrm{M}_{\odot}$ with dimensionless component spins of up to $0.9895$.
Injections are made at $D_{\mathrm{eff}}$ up to and beyond the horizon distance \cite{Abadie:2012lm}, the distance at which an optimally located and oriented binary would produce an expected signal-to-noise ratio of 8, of the detectors so that the signal recovery of MBTA can be tested.
The result from this investigation can be seen in figure\,\ref{fig:missedfound}, where $\mathcal{M}$ is plotted against $D_{\mathrm{eff}}$ for the NSBH injections.
MBTA recovers injections with better than $50\%$ efficiency up to $\mathtt{\sim}400$\,Mpc.

\subsection{Signal consistency test}
\label{sec:mfo}

The matched filter time-series signal consistency test gives improved separation of signal and noise events based on their shape in the matched filter time-series of each detector.
Signal events cause a single narrow feature in the matched filter time-series, while noise events produce wider features with multiple maximums.
The matched filter time-series signal consistency test has two tunable parameters as seen in equation\,\ref{eq:mfo}, $A$ and $B$.
To empirically find the best values for these parameters, the matched filter time-series signal consistency test was tuned on a set of BNS injections to obtain good background rejection.
These tests gave the ``standard'' values of $A = 65$ and $B = 0.4$.
Using these values the matched filter time-series signal consistency test was then tested on a set of NSBH injections.
Comparing the standard matched filter time-series signal consistency test to a number of other tunings, the standard tuning was found to be the best compromise of performance for both BNS and NSBH.

\begin{figure*}
  \captionsetup[subfigure]{labelformat=empty}
  \centering
    \subfloat{\includegraphics[width=0.7\textwidth]{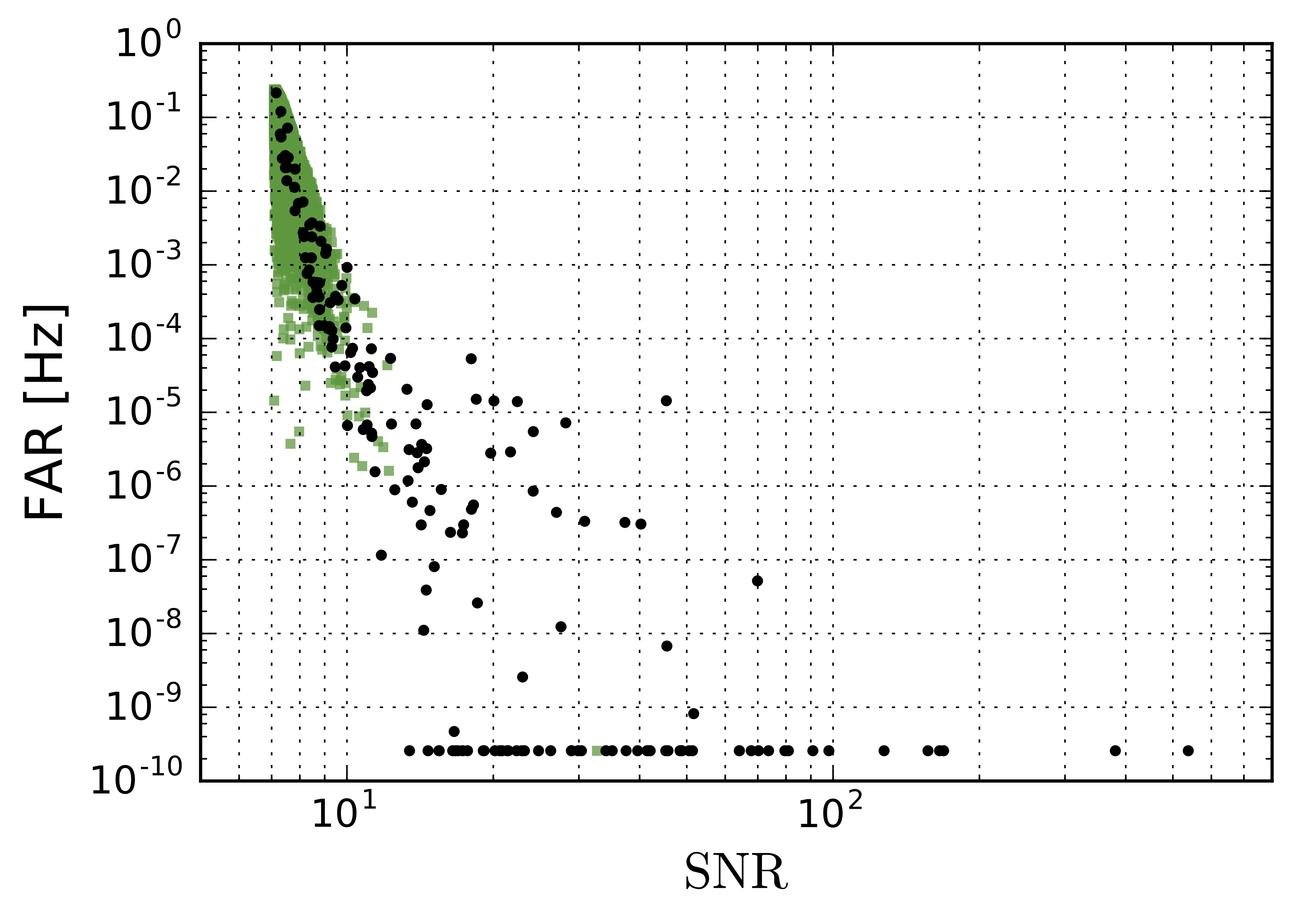}
    }\\
    \subfloat{\includegraphics[width=0.7\textwidth]{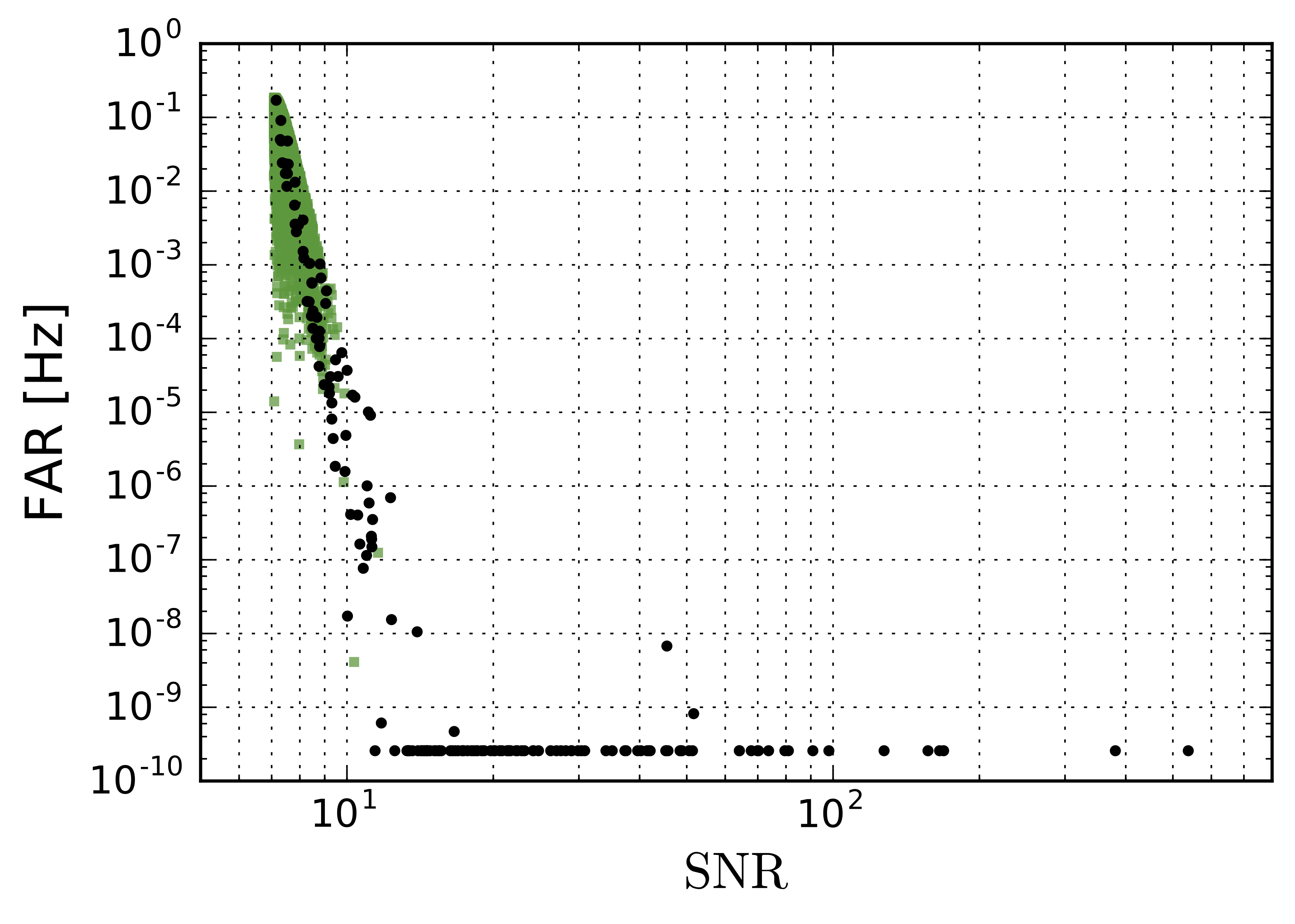}
    }
  \caption{\label{fig:mfocomparison}FAR against SNR for the ``NSBH MDC'' without the matched filter time-series signal consistency test (top) and with the standard matched filter time-series signal consistency test (bottom).
           Background noise events are shown as light green squares, while injections events are shown as black circles.}
\end{figure*}

Without the matched filter time-series signal consistency test, high SNR noise events contaminate the background used to compute the FAR for the injection events, which increases the FAR assigned to higher-SNR injections.
The matched filter time-series signal consistency test removes high SNR noise events, thus reducing the FARs computed for louder injections.
This result is shown in figure\,\ref{fig:mfocomparison} for the ``NSBH MDC'', as outline in section\,\ref{sec:signalrecovery}.

\subsection{Exact match coincidence test and improved sky localisation}
\label{sec:exactmatch}

In preparation for the advanced detector era, Bayestar, the rapid Bayesian position reconstruction code that produces sky maps for MBTA triggers, was upgraded to perform improved sky localisation and now requires phase information as well as the usual amplitude and timing for events.
To account for this the exact match coincidence test, which requires that an event is detected in all instruments with the same virtual template (same component masses and spins), was implemented in MBTA so that the relative phase across detectors for a signal can be accurately reported.

To validate the Bayestar sky localisation using MBTA events, the ``sky localisation MDC'' which was previously analysed with the GSTLAL pipeline \cite{Singer:2014iq}, was analysed with MBTA.
This MDC was generated to predict the detection, sky localisation, and parameter-estimation capabilities of the LIGO detectors during the first observing period of the advanced detector era, O1.
An equivalent analysis was performed with MBTA and compared to the previous results to verify that consistent sky localisations are obtained.

For this MDC, Gaussian noise coloured to match the expected PSD of the early Advanced LIGO detectors during O1 was used.
This MDC contains $3.5$\,PN SpinTaylorT4 BNS injections with component masses of $1.2\,\mathrm{M}_{\odot}$ to $1.6\,\mathrm{M}_{\odot}$ and dimensionless component spins of up to $0.05$.

\begin{figure*}
  \captionsetup[subfigure]{labelformat=empty}
  \centering
    \subfloat{\includegraphics[width=0.5\textwidth]{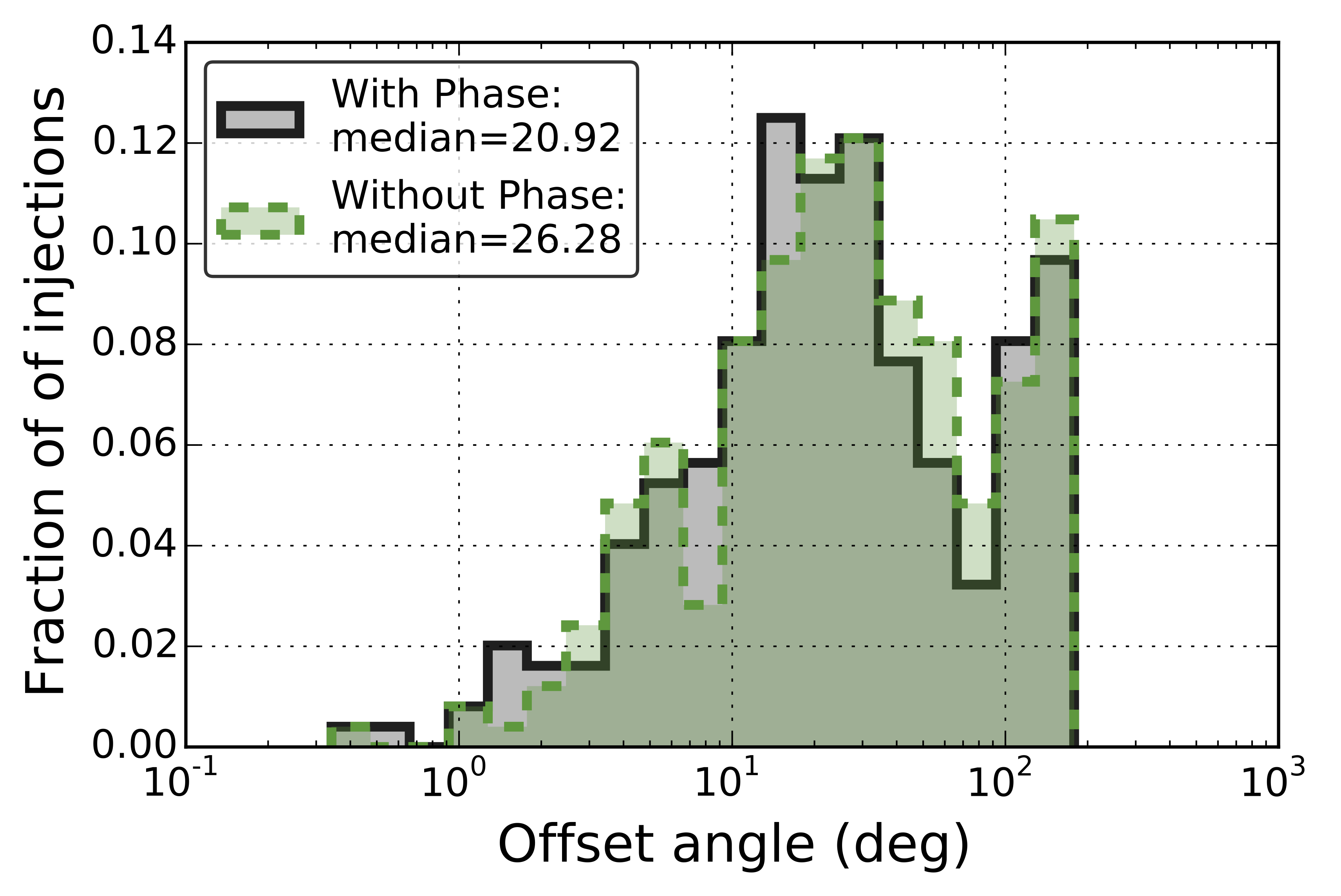}
    }
    \subfloat{\includegraphics[width=0.5\textwidth]{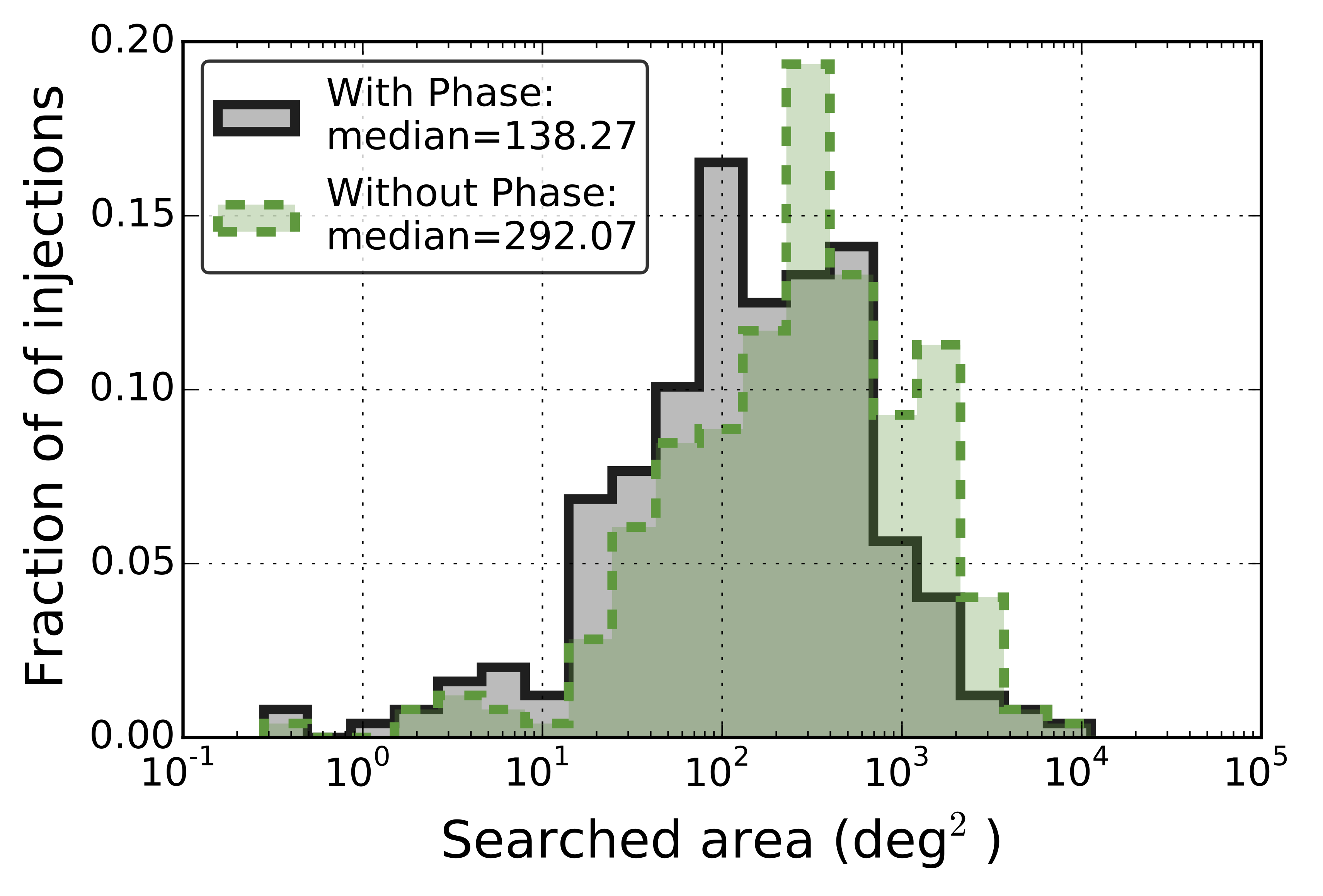}
    }
  \caption{\label{fig:phase}Offset angle histogram (left) and searched area histogram (right) for the Bayestar sky localisations performed on simulated GW events recovered by MBTA in the sky localisation MDC.
           Results from the sky localisations performed without using phase information are shown in light green, while sky localisations performed using the phase information are shown in black.}
\end{figure*}

To test the improved sky localisation provided by the recently upgraded Bayestar, sky localisations for simulated GW injections recovered from the sky localisation MDC by MBTA was performed.
These were processed twice with Bayestar, first using just timing and amplitude information, and a second time also utilising the phase information allowing the improved sky localisation.
In figure\,\ref{fig:phase} an improvement in the offset angle, which is defined as the angle between the sky location of the injected GW signal and the mode of the posterior, can be seen when utilising the phase information for the sky localisation.
There is also an improvement in the searched area, which is defined as the area of the highest confidence region around the mode of the posterior to include the sky location of the injected GW signal.

Comparing the sky localisation performance of MBTA to that obtained with GSTLAL, both pipelines were seen to perform similarly, with the median of the offset angle and searched area distributions agreeing within a few percent.
This agrees with what is expected as the pipelines recover signals using different methods and provide slightly different amplitude, timing, and phase information.

\section{Running in the advanced detector era}
\label{sec:adv_era}

During the advanced detector era, the MBTA pipeline will be run online during each of the observing periods.
To maximise the observation time, the maintenance of the online pipeline will be co-ordinated with detector downtime which will increase the likelihood of a GW detection.
The BNS and NSBH template bank, as described in section\,\ref{sec:signalrecovery}, will be used in O1 for the online search with MBTA.
This template bank covers the sources that are expected to produce the brightest EM counterparts.

The filtering, which is the most computationally demanding part of the pipeline, can be run in real time on a single 32-core machine (Xeon E5-2650 @ $2.60$\,GHz) for a single detector's data using the BNS and NSBH template bank of approximately $200,000$ templates for O1.
Using the O1 configuration the latency of MBTA from receiving the data to event production is $\mathtt{\sim}30$\,s.

To monitor the status of the pipeline online, a rota of pipeline experts is set up to monitor the pipeline throughout the observing periods.
Although the MBTA pipeline is very stable when running online, the regular monitor tasks mainly involve checking the status of the pipeline processes, monitoring data transfer, and following the status of each of the detectors and commissioning activities.
It is also the responsibility of the monitor to check a number of pipeline figures of merit for unusual behaviour, such as outliers in the cumulative FAR distribution above what is expected from regular statistics, excesses in the single detector trigger rates, or variations in the data transfer latency, and then respond accordingly.
It is also the duty of the monitor to be the MBTA contact with detector and EM follow-up experts for any interesting events identified by MBTA.
This includes the follow-up of GW event candidates, which are subjected to a number of validation tests before release to astronomical partners for EM follow-up.

During the first observing periods, there is expected to be very little in the way of online data quality flags.
Therefore, as well as the $\chi^{2}$ cut and matched filter time-series signal consistency test, other techniques can also be considered to reduce the impact of noise on the analysis.
One option is to perform automated self vetoing on periods of loud detector noise.
When there is a large SNR noise event in the data stream, it makes the analysis blind to astrophysical signals and causes a residual reduction in the pipeline sensitivity for a few hundred seconds due to the effect of the noise on the PSD estimation.
This technique can be applied when there is a sudden large drop in the BNS range, caused by instrumental noise, to remove this noisy data from the PSD estimation and so retain sensitivity throughout periods of instrumental noise.

The MBTA pipeline is now ready for the detection of GW signals from CBCs which are of interest for EM follow-up with sub-minute latency, and the exciting prospects the advanced detector era will bring.

\section*{Acknowledgements}

We would like to thank the CBC group for useful discussions and assistance with MDC data preparation, Ian Harry for guidance in using the LALsuite template bank generation tools, and Tito Dal Canton for assistance with the optimal SNR computation.
We also thank Leo Singer for guidance in running Bayestar using MBTA triggers and assistance with setting up the sky localisation MDC, Riccardo Sturani for discussions on waveform generation, and Thomas Dent for his work on pipeline comparison studies.
Finally we would like to thank the MBTA review team, Gijs Nelemans and John Whelan, for a constructive review and discussions.
TA has received support from Labex ENIGMASS.

\section*{References}
\bibliography{P1500245.bbl}

\end{document}